\documentclass[fleqn,twoside,a4]{article}
\usepackage{espcrc2}
\usepackage{amsmath}

\usepackage{graphicx}
\usepackage[figuresright]{rotating}

\renewcommand{\thefootnote}{\fnsymbol{footnote}}

\title{A comparative study of overlap and staggered fermions in the Schwinger
model$^*$}

\author{Stephan D\"urr\address{DESY Zeuthen, Platanenallee 6,
    D-15738 Zeuthen, Germany}
  and
  Christian Hoelbling\address{Centre de Physique Th\'eorique$^\dag$,
    CNRS Luminy, F-13288 Marseille Cedex 9, France}}

\begin{document}



\begin{abstract}
We investigate the validity of the square rooting procedure of the staggered
determinant in the context of the Schwinger model. We find some evidence that
at fixed physical quark mass the square root of the staggered determinant
becomes proportional to the overlap determinant in the continuum limit. We
also find that at fixed lattice spacing moderate smearing dramatically
improves the chiral behavior of staggered fermions.
\vspace{-4mm}
\end{abstract}

\maketitle

\footnotetext[1]{Talk presented by Christian Hoelbling at Lattice 2004.}
\footnotetext[2]{Unit{\'e} Mixte de Recherche (UMR 6207) du CNRS et des
Universit{\'e}s Aix Marseille 1, Aix Marseille 2 et sud Toulon-Var,
affili{\'e}e {\`a} la FRUMAM.}

\renewcommand{\thefootnote}{\arabic{footnote}}

\vspace{-1mm}
\section{Introduction}
\vspace{-2mm}
Recently, unquenched lattice QCD calculations with improved staggered
fermions have had remarkable success in reproducing a variety of
phenomenologically interesting quantities \cite{greatsuccess}. In these
calculations, the determinant of a single fermion is obtained as the fourth
root of the staggered determinant. It is a priori unclear whether there
exists a local operator $D$ describing a single fermion flavor
with
\begin{equation}
\label{eq.locop}
\text{det}\left(D\right)\propto
\text{det}\left(D_{\text{stag}}\right)^{1/4}
\end{equation}
and it has been shown \cite{Bunk:2004br} that a naive guess leads to a
nonlocal operator. Furthermore, the lack of an (exact) index theorem for
staggered fermions raises the question how far into the chiral regime one can
push calculations at fixed lattice spacing.

Given this situation, we decided to study the behavior of staggered fermions
in the simple and well known Schwinger model (QED$_2$). We focus on
observables that - like the determinant - are obtainable from the Dirac
spectrum alone. As a point of reference, we compare the staggered data to
results obtained with overlap fermions, which are known to be free of
conceptual problems. We also investigate UV-filtered staggered and overlap
fermions, which are technically realized by APE-smearing the gauge
backgrounds. For the U(1) gauge group smearing consists of
taking a weighted average of the phases of the original link and the
staple. Due to the twofold fermion doubling in 2D, the $N_f=1$ case is
obtained by weighting the gauge configurations with the square root of the
staggered determinant. Technical details of our simulations can be found in
\cite{Durr:2003xs}, related studies in QCD are reported in
\cite{Durr:2004xx,Wong:2004ai}.

\vspace{-1mm}
\section{Chiral condensate}
\vspace{-2mm}
Due to the Mermin-Wagner theorem, the chiral condensate of the continuum
massless Schwinger model vanishes for $N_f\ge 2$
\cite{MerminWagnerHohenbergColeman}. For $N_f=1$ the formation of a
condensate does not imply spontaneous symmetry breaking. In the massless
case, its infinite volume value is analytically known \cite{Schwinger:tp}
\begin{equation}
{\langle\bar\psi\psi\rangle\over e}=
{e^\gamma\over 2\pi^{3/2}}\simeq 0.1599...\;\;.
\end{equation}
Away from the chiral limit one expects an additional
contribution to the condensate proportional to the quark mass and at
most logarithmically divergent in the cutoff. Also, due to the exact/remnant
chiral symmetry of overlap/staggered fermions, the chiral condensate
renormalizes multiplicatively and due to the dimensionful coupling
$[e]=[a]^{-1}$ all renormalization factors are $Z=1+O(a^2 e^2)$.

\begin{figure}[t]
\centerline{
\includegraphics[width=18pc]{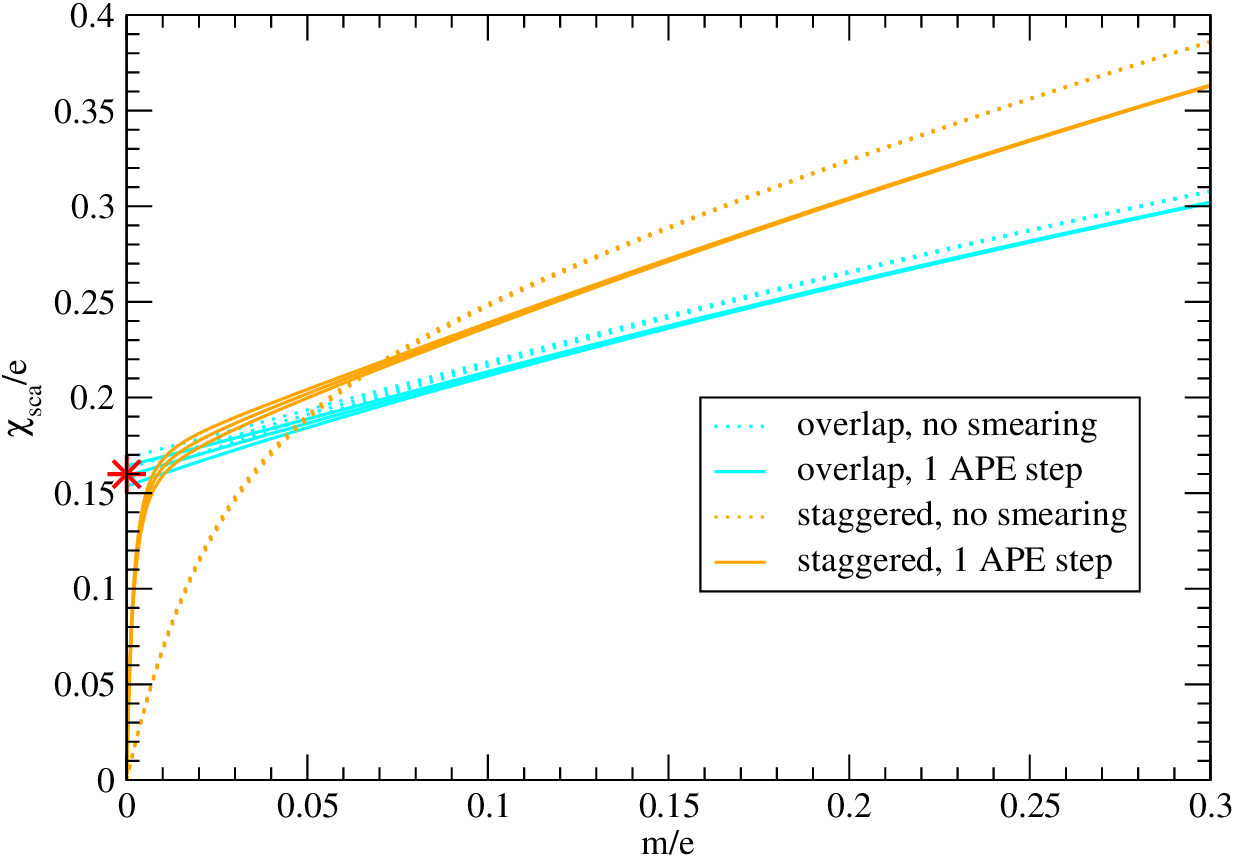}
}
\vspace{-2pc}
\caption{$\langle\bar\psi\psi\rangle/e$ of staggered and overlap fermions with
$N_f=1$ on a $24^2$ lattice
at $\beta=7.2$. The star denotes the analytically known value of the continuum chiral condensate
from \cite{Schwinger:tp}.}
\label{fig.chcond}
\end{figure}

In fig.~\ref{fig.chcond} we plot the $N_f=1$ chiral condensate versus the quark
mass, as obtained from staggered and overlap fermions. Our results for the
unfiltered operators (dotted lines) indicate that
overlap fermions show a nice chiral behavior while staggered
fermions behave qualitatively wrong in the chiral limit.
By contrast, the UV-filtered staggered fermions (full line) exhibit a much better
behavior down to comparatively small quark masses. They still show, ultimately,
a qualitatively wrong behavior in the chiral limit, but the mass at which
it sets in is drastically reduced.


\vspace{-1mm}
\section{Spectral mimicry}
\vspace{-2mm}
A finite condensate in the chiral limit is related to the existence of
exact zero modes. The dramatically improved behavior of
$\langle\bar\psi\psi\rangle/e$ for UV-filtered staggered fermions
down to fairly small quark masses therefore suggests that they develop near-zero
modes on topologically nontrivial gauge
configurations and that very small quark masses are needed to expose these modes
as not being true zero modes (see \cite{Durr:2003xs} for a more comprehensive
discussion of this point). It is thus interesting to directly compare
staggered and overlap spectra on individual configurations.

\begin{figure}[t]
\centerline{\includegraphics[width=18pc]{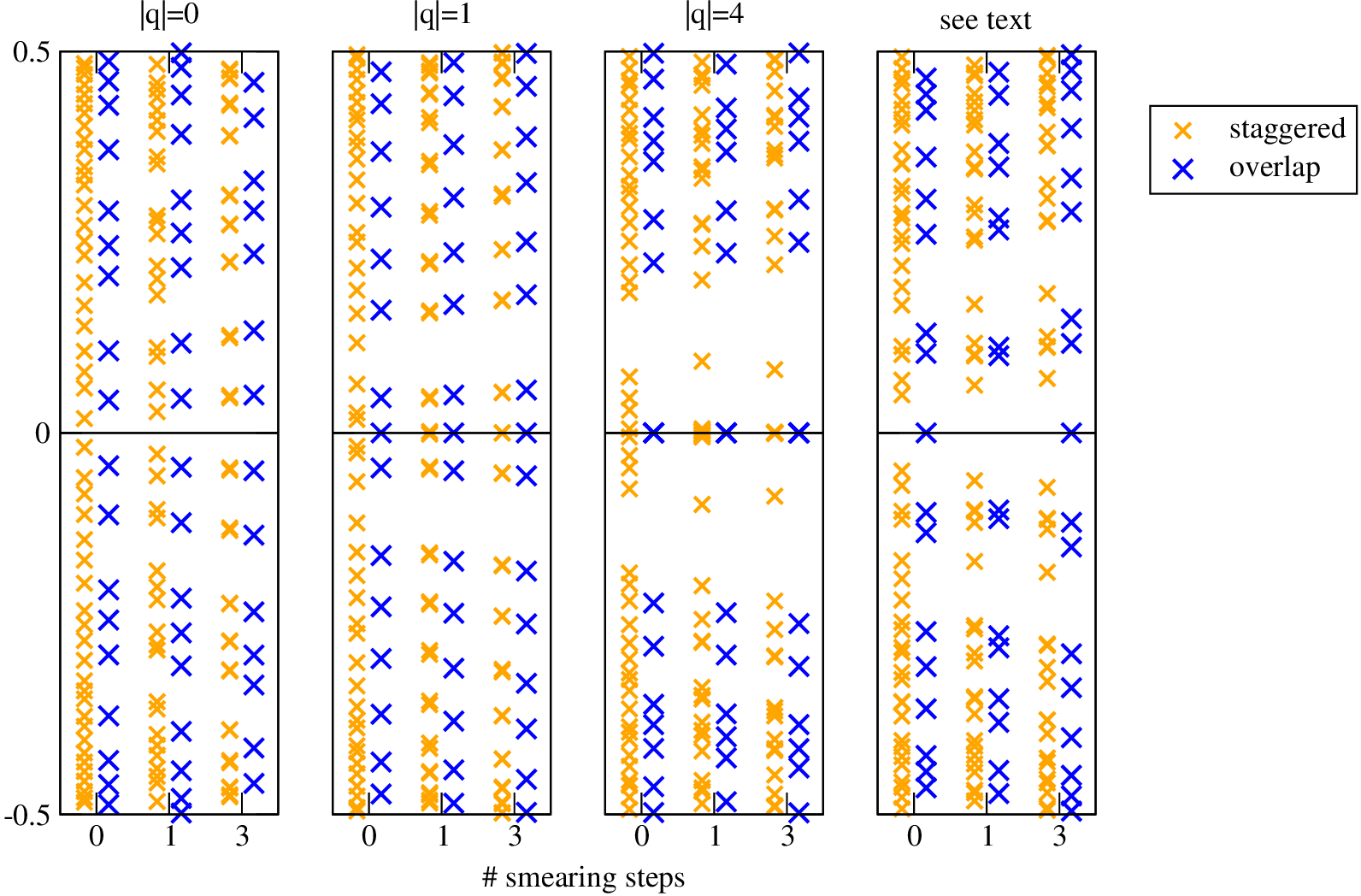}}
\vspace{-2pc}
\caption{Comparison of staggered versus overlap spectra for three typical
configurations ($3$ left columns) and one configuration with undecided
topology (right column).}
\label{fig.spec}
\end{figure}

Fig.~\ref{fig.spec} shows a comparison of the IR part of massless staggered and
overlap\footnote{For the overlap operator, we plot the (purely imaginary)
eigenvalues $\hat\lambda=(1/\lambda-1/2\rho)^{-1}$ of the chirally rotated
operator (see \cite{Durr:2003xs} for more details).} spectra on some selected
configurations on a $20^2$ lattice at $\beta=4$.

The two plots on the left show typical configurations with topological charge
$q=0$ and $|q|=1$, respectively. The unfiltered staggered spectra do not
resemble the overlap spectra and it is difficult to see the qualitative
difference between $q=0$ and $|q|=1$. After moderate UV-filtering however, the
eigenvalues of the staggered operator form near-degenerate pairs
which sit close to a single overlap eigenvalue. In particular, in the
$|q|\!=\!1$ case a pair of eigenmodes moves very close to the real axis,
confirming the expectation about the existence of near-zero modes that we had
from the chiral condensate analysis.

The third column of fig.~\ref{fig.spec} shows a typical configuration with
higher topological charge ($|q|\!=\!4$). Here, after $3$ smearing steps only
$3$ pairs of eigenmodes have managed to come close to the real axis and the
fourth one is still further out.

The last column of fig.~\ref{fig.spec} presents a selected ``worst case''
configuration on which the topological charge varies repeatedly under
subsequent smearing steps. Qualitative resemblance between the two types of
spectra is vague at best and there is no evidence for a pair of
staggered eigenvalues moving close to the real axis.
Such configurations are rare already at $\beta\!=\!4.0$.

\vspace{-1mm}
\section{Determinant ratio}
\vspace{-2mm}
The qualitative similarity between the IR part of staggered and overlap
spectra on individual configurations leads to an interesting suggestion
regarding the local operator $D$ in (\ref{eq.locop}). If
eigenvalue pairs become truly degenerate in the continuum limit and
topologically undecided configurations are are strictly O$(a^2)$ effects, it
could be that $D_{\text{overlap}}$ is a local operator with
\begin{equation}
\label{eq.locop2d}
\text{det}\left(D_{\text{stag}}\right)^{1/2}
\propto
\text{det}\left(D_{\text{overlap}}\right)+O(a^2)
\;.
\end{equation}

\begin{figure}
\includegraphics[height=18pc,angle=270]{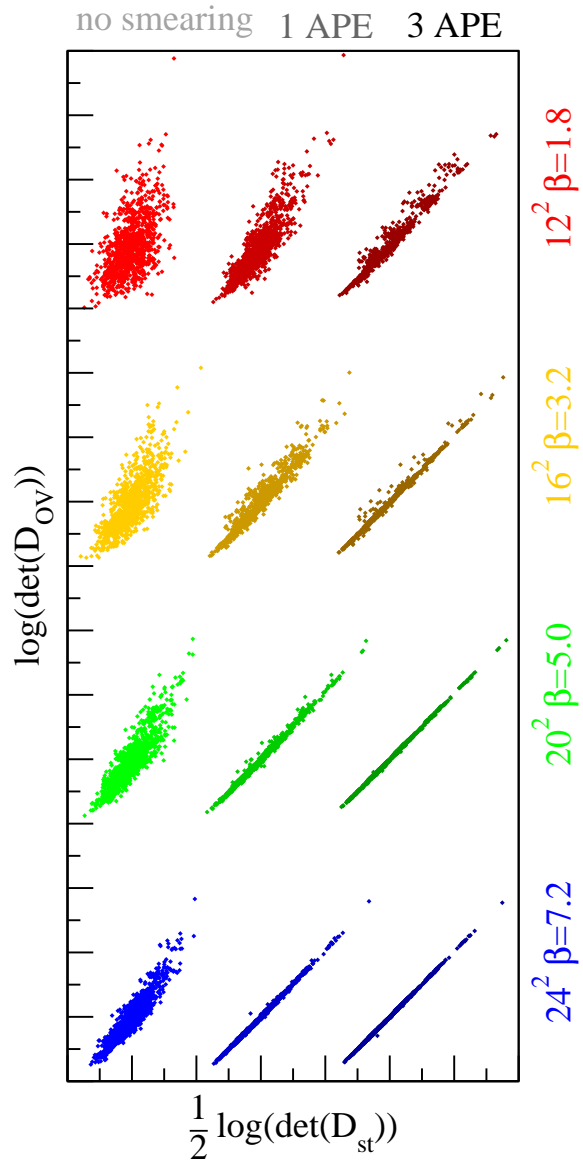}
\vspace{-2pc}
\caption{Log-Log scatter plot of the overlap determinant versus the square
root of the staggered determinant at 4 lattice spacings and 3 UV-filtering
levels and fixed fermion mass $m/e\!=\!0.035$. Every plot shows 1000
decorrelated quenched configurations at fixed physical volume. The
individual plots are offset for better visibility. In all cases the slope is
consistent with $1$.}
\label{fig.detrat}
\end{figure}

Fig.~\ref{fig.detrat} shows a scatter plot of
$\text{log}(\text{det}(D_{\text{stag}})^{1/2})$
vs. $\text{log}(\text{det}(D_{\text{overlap}}))$ for different lattice
spacings and UV-filtering levels at one fixed physical quark mass. One can
clearly see that there is a tendency for the determinants to be proportional
in the continuum limit and that the proportionality is better developed for
the UV-filtered operators.

With a careful scaling study one might be able to test
(\ref{eq.locop2d}). The real goal is, of course, to see whether such a
relation holds in QCD. Preliminary studies in this direction can be
found in \cite{Durr:2004xx}.

\bigskip
{\bf Acknowledgments:}
S.D.\ is supported by DFG,
C.H.\ by EU grant HPMF-CT-2001-01468.

\end{document}